# Space-Time Modulated vs 2-Bit Reflect-Arrays: A Comparison of Main Beam Gain and Sidelobe Level Performance

Muhammad Rizwan Akram, *Member, IEEE*, and Douglas H. Werner, *Fellow, IEEE*

***Abstract*— This paper presents a comprehensive analysis of the phase errors introduced by 2-bit reflectarray architectures and investigates their impact on aperture synthesis for beam-steering applications. Building upon this analysis, an amplitude-aware synthesis technique based on space-time modulated (STM) reflect-arrays is proposed to mitigate the main-beam gain degradation commonly associated with STM-based equivalent phase generation while preserving the enhanced phase resolution enabled by temporal coding. A $10\lambda \times 10\lambda$ reflect-array aperture with $\lambda/2$ inter-element spacing is considered, where the desired aperture phase distribution is synthesized using a library of achievable reflection coefficients generated from all possible combinations of four 2-bit phase states distributed across ($L$) temporal segments. The proposed approach introduces a phase-tolerance window and selects the reflection state with the highest reflectance within the allowable phase range. By exploiting the inherent amplitude diversity of the STM state space, the proposed approach maximizes aperture efficiency while maintaining the desired beam-steering phase distribution. Numerical results demonstrate that the method substantially reduces the gain penalty typically associated with STM reflect-arrays along with providing significant sidelobe suppression.**

## I. Introduction

Electronically reconfigurable reflect-arrays [1-3] have attracted considerable attention as a promising solution for realizing high-gain beamforming apertures with reduced hardware complexity compared to conventional phased arrays [4, 5]. By introducing tunable phase control at each reflecting element, active reflect-arrays can dynamically manipulate the aperture phase distribution to achieve beam steering [6], beam shaping [7], and multibeam operation [8] while preserving the low-profile and lightweight characteristics desirable for satellite communication platforms. Such capabilities are particularly attractive for emerging multi-orbit satellite networks and spaceborne communication systems [9-11], where electronically agile beams are required to support dynamic coverage [12], user tracking [13, 14], and adaptive resource allocation [15].

Consequently, significant research efforts have focused on the development of active reflect-array architectures employing varactor diodes [16, 17], PIN diodes [18-22], liquid-crystal technologies [23], and other tunable mechanisms to enable real-time reconfiguration [24, 25]. Continuous phase tuning approaches, typically based on varactor diodes or tunable materials, are attractive due to their ability to provide near-continuous phase compensation and fine beam steering resolution. However, these solutions often provide limited phase control and typically suffer from higher losses due to high intrinsic resistance [16, 23]. Conversely, switching-based architectures employ PIN diodes or RF switches to realize discrete phase states with fast switching speeds, and full phase range coverage. Nevertheless, phase quantization introduces aperture errors that reduce beamforming efficiency and increase sidelobe levels. It is well established that 1-bit, 2-bit, and 3-bit phase quantization incur quantization losses of approximately 3 dB, 0.9 dB, and 0.2 dB, respectively [26, 27]. While higher-bit implementations can approach the performance of continuously tunable reflect-arrays, they require a substantially larger number of switching elements, bias lines, and control circuitry, thereby increasing cost, insertion loss, and implementation complexity. Consequently, 2-bit phase quantization [22, 26] has emerged as a particularly attractive compromise, providing four discrete phase states that achieve near-ideal beamforming performance with relatively modest hardware complexity, making it a practical as well as scalable solution for large-aperture electronically steerable reflect-arrays.

Despite its favorable complexity-performance tradeoff, phase quantization remains a fundamental limitation in digitally controlled reflect-arrays. In an ideal reflect-array, the aperture phase distribution is continuously tailored to compensate for the spatial phase delay associated with the feed illumination while simultaneously imposing the desired beamforming phase gradient. In practical 2-bit implementations, however, the required continuous phase profile must be approximated using only four discrete phase states, resulting in unavoidable quantization errors across the aperture. These phase errors introduce deviations from the ideal wavefront, leading to a reduction in aperture efficiency and main beam pointing inaccuracies while increasing sidelobe levels and undesired

This work was supported in part by the CesiumAstro under contract no. 303314. The manuscript was initially submitted on July 14, 2026. The authors are affiliated with School of Electrical Engineering and Computer Science, The Pennsylvania State University, University Park, PA 16802, USA. Corresponding Author: Muhammad Rizwan Akram (e-mail: rizwanakram373@ gmail.com).

radiation in non-target directions. The impact becomes particularly pronounced for large steering angles and complex beamforming scenarios, where the spatial phase gradient varies rapidly across the aperture and the discrepancy between the desired and available phase states increases.

To alleviate the performance degradation associated with low-bit phase quantization, spatiotemporally modulated (STM) reflect-arrays and coding metasurfaces [28-31] have recently emerged as a promising alternative to conventional 2-bit architectures. By periodically switching the reflection phase states according to a prescribed temporal coding sequence, STM approaches exploit the additional temporal degree of freedom to synthesize effective complex reflection coefficients [31] with substantially finer phase and amplitude resolution than those available from the underlying hardware states. Consequently, the achievable aperture distribution can more closely approximate the desired continuous phase profile, leading to reduced quantization errors, suppressed sidelobe levels, and improved beamforming fidelity. Despite these advantages, the realization of multi-level equivalent phase states often relies on coding sequences with nonuniform duty cycles, resulting in effective amplitude variations across the aperture that further decrease the available power in the main beam [31]. As a result, STM-based reflect-arrays generally exhibit a tradeoff between sidelobe suppression and reduced gain of the main beam.

This paper addresses the reduction in main beam gain commonly associated with space-time modulation while simultaneously maintaining significantly lower sidelobe levels. The proposed approach employs an optimization strategy that prioritizes high-amplitude states and selects phase values within the allowable phase-tolerance window for each reflect-array element. As a result, the synthesized amplitude and phase distribution achieves higher main beam efficiency and improved sidelobe suppression compared to previously reported 3-bit and 4-bit equivalent space-time modulation synthesis approaches [31].

The remainder of this paper is organized as follows. Section I analyzes the impact of phase quantization in a conventional 2-bit reflect-array by examining the gap between the desired and available phase states. Section II presents the theoretical foundations of space-time modulation. Section III describes the proposed optimization-based synthesis approach. Section IV presents the numerical results and discussion, followed by the conclusions.

## II. Quantization Impact

In a conventional 2-bit reflect-array, the available reflection phases are restricted to four discrete states separated by 90°, typically corresponding to 0°, 90°, 180°, and 270°. For a desired beam-steering phase distribution, the required continuous phase at each element must therefore be approximated by the nearest available discrete state, introducing phase quantization errors. The maximum phase error is bounded by ±45° as shown for two different scenarios in Fig. 1, while the average error across the aperture results in a reduction of coherent field superposition in the desired beam direction. Consequently, a portion of the radiated power is redistributed from the main beam into undesired directions, leading to reduced aperture efficiency, lower realized gain, and elevated sidelobe levels as can be clearly observed in Fig. 2 for two different steering angles. The severity of these effects increases for large apertures and wide-angle beam steering, where the required phase gradient varies rapidly across the surface. Although 2-bit architectures offer a favorable trade-off between hardware complexity, biasing requirements, and performance, the inherent mismatch between the desired continuous phase distribution and the available discrete phase states results in radiation in unintended directions. This limitation motivates the development of advanced synthesis techniques capable of generating a larger set of effective phase states while maintaining the simplicity of a 2-bit hardware implementation.

## III. Space Time Modulation Theory and Design

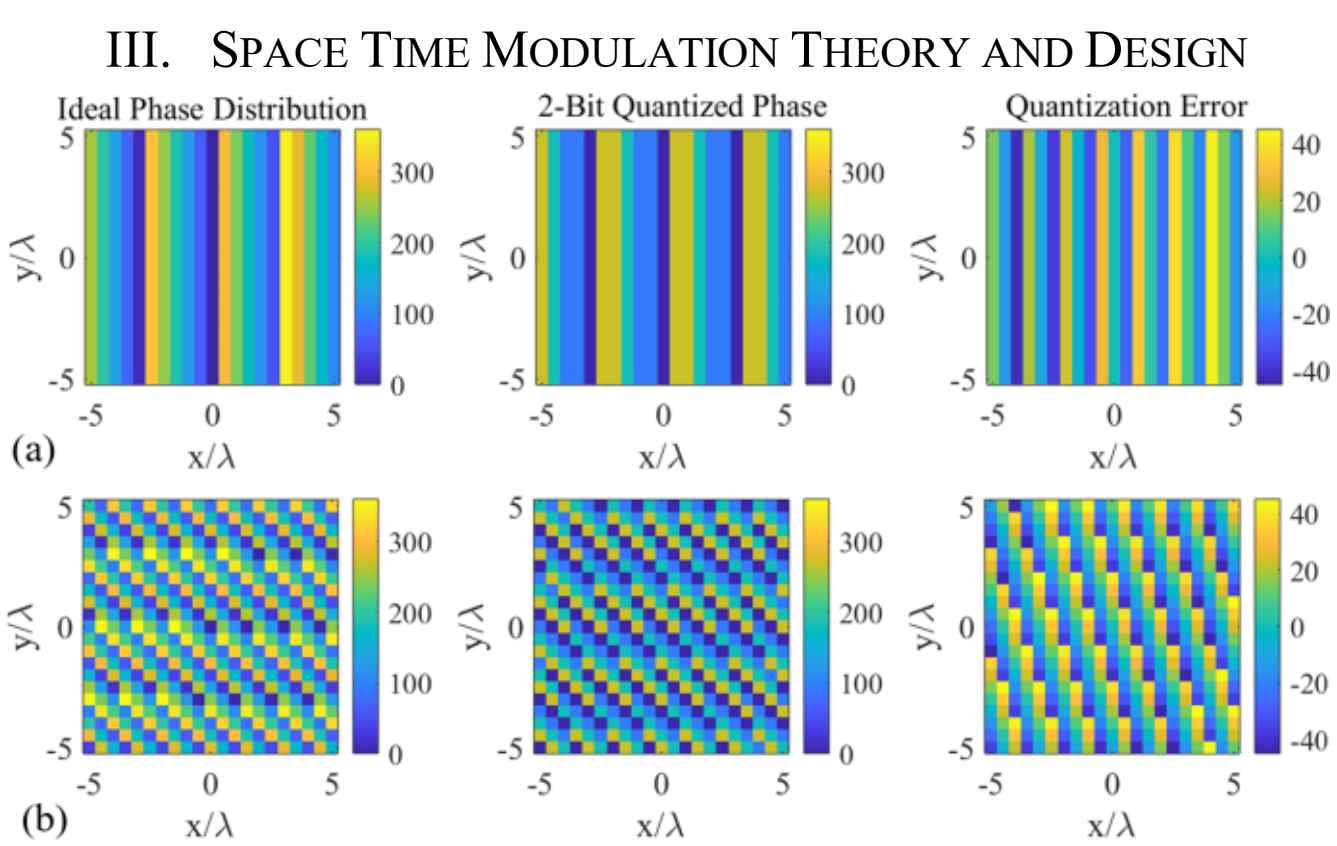

Fig. 1. Aperture phase distribution and deviation from ideal phase over an aperture having a size of 10λ × 10λ with inter-element spacing of d = λ/2 and beam steering angles of (a) $\theta$ = 20º, $\phi$ = 0º and (b) $\theta$ = 60º, $\phi$ = 40º.

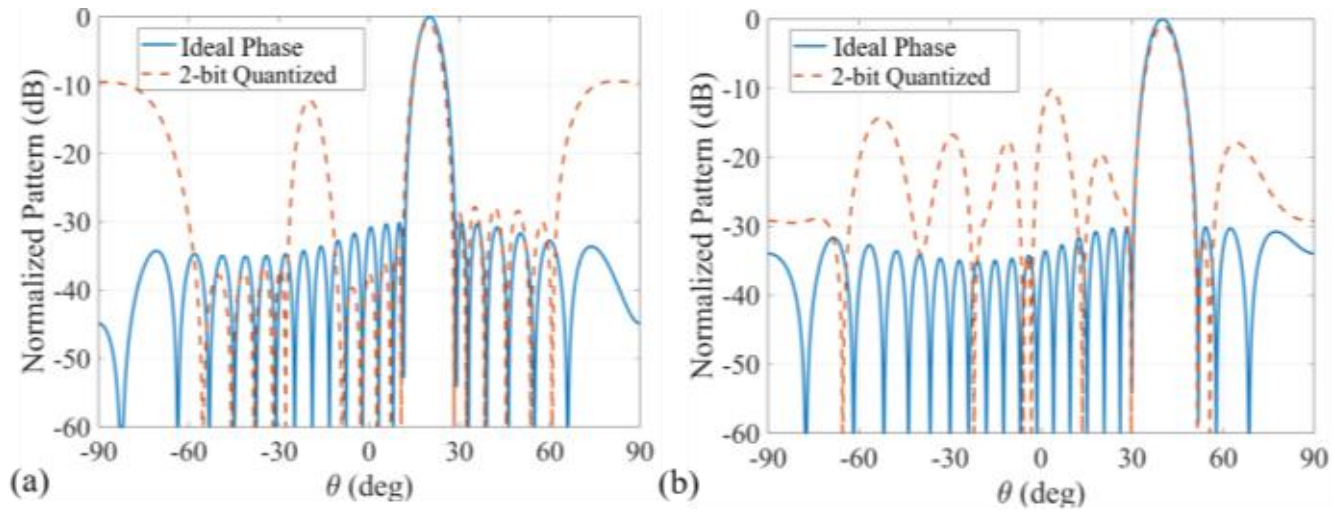

Fig. 2. Normalized radiation pattern from an aperture having a size of 10λ × 10λ with inter-element spacing of d = λ/2 and beam steering angles of (a) $\theta$ = 20º, $\phi$ = 0º and (b) $\theta$ = 40º, $\phi$ = 0º.

We will start by considering a reflect-array unit cell capable of providing 2-bit phase control with unity amplitude. By applying different control voltages to the PIN diodes, the reflection phase can be switched from 0º, 90º, 180º, 270º, enabling the two-bit phase control while providing unity amplitude. The reflection coefficient is represented by $\Gamma_q = \{1,\ j, -j, -1\}$ for the 2-bit reflect-array and is modulated by a control pulse with a period of $T_0$. The period can be further subdivided into $L$ segments to enable another degree of freedom through quantized time sequences with each time segment given by $\tau = \frac{T_0}{L}$. Hence, the reflection coefficient during each segment ($l$) can be represented by $\Gamma(\mathrm{t}) = \Gamma_l$ for $l\tau < t < (l+1)\tau$ and $\Gamma_l \in \Gamma_q$. Now the periodic reflection coefficient can be represented by using a Fourier series as $\Gamma(\mathrm{t}) = \sum_{m=-\infty}^{\infty} \mathrm{a_m} e^{j\omega_0 m t}$ where $\omega_0 = \frac{2\pi}{T_0}$. By applying Fourier transform theory, we can evaluate the harmonics generated as $a_m = \frac{1}{T_0}\int_0^{T_0}\Gamma(t)e^{-j\omega_0 mt}dt$. Since the focus of the current work is on the gain enhancement at the carrier frequency, the reflection response at the carrier frequency can be represented as $a_0 = \frac{1}{T_0}\int_0^{T_0}\Gamma(t)dt = \frac{1}{L}\sum_{l=0}^{L}\Gamma_l$. This expression can be utilized to synthesize an arbitrary reflection phase by selecting different combinations of available 2-bit phase values within $L$ quantized time segments in each pulse of period $T_0$.

For example, to synthesize a phase value of $45^o$ using time segments of $L = 4$, $a_0 = \frac{1}{4}(1+1+j+j) = 0.707\angle 45^o$. Similarly, a phase value of $18^o$ using time segments of $L$= 8 can be synthesized as $a_0 = \frac{1}{8}(6+2j) = 0.79\angle 18.4^o$. Now to generalize the formula, let's assume the occurrence of the phase states are represented as $n_q$where $q$ represents four different states of the 2-bit phase control of the reflect-array unit cell. Now the reflection response at the carrier frequency can be obtained as $a_0 = \frac{1}{L}(n_1 - n_3 + j(n_2 - n_4))$. The equivalent amplitude and phase of the reflection response are found to be $|a_0| = \frac{1}{L}\sqrt{(n_1-n_3)^2 + (n_2-n_4)\text{^}2}$, $\angle \tan^{-1}\left(\frac{n_2-n_4}{n_1-n_3}\right)$ respectively.

Following the STM formulation described in the previous section, a library of achievable reflection coefficients is generated for $L$ control segments. The library is constructed by enumerating all possible occurrences of the four available 2-bit phase states within the ($L$) time segments. Since only the number of occurrences of each phase state contributes to the available reflection states, all valid non-negative integer combinations satisfying $n_1 + n_2 + n_3 + n_4 = L$ are considered. The total number of achievable STM states therefore can be given by $N_{STM} = \binom{L+3}{3}$ which evaluates to 35, 165 and 969 for $L$ = 4, 8 and 16 respectively.

Once the library for STM possible reflection states is generated, the next step is to evaluate the desired phase at the 2D aperture. Let's assume a reflect-array aperture of $10\lambda \times 10\lambda$ consisting of subwavelength elements arranged on a uniform lattice with an inter-element spacing of $\lambda/2$. For a desired beam steering direction $(\theta, \phi)$, the required ideal phase distribution is first calculated using conventional reflect-array phase synthesis. This phase profile serves as the target aperture distribution and represents the ideal phase response required at each reflect-array element to form the desired far-field scattering pattern.

The desired phase at each aperture element is compared with all available phase states contained in the library. Instead of selecting the state that minimizes phase error alone, a phase tolerance window $\Delta\phi$ is introduced around the desired phase. All STM states whose phase values fall within this tolerance range are identified as valid candidates. Among these candidates, the state exhibiting a reflection amplitude close to unity is then selected and assigned to the corresponding reflectarray element. Consequently, the proposed synthesis procedure transforms the aperture design problem into an amplitude-aware phase-matching process, where phase accuracy is maintained while maximizing reflection amplitude.

## IV. Results and Discussion

The synthesized amplitude and phase map over an aperture of $10\lambda \times 10\lambda$ with inter-element spacing of $\lambda/2$ is shown in Fig. 3 for two different steering angles, *i.e.* θ = 20º and θ = 40º, assuming a phase tolerance of $\Delta\phi = 15^o$. The proposed synthesis approach maximizes the utilization of unity-amplitude states while selecting phase states within the prescribed tolerance window. As a result, average aperture amplitudes of 0.85 and 0.88 are achieved, with minimum amplitudes of 0.73 and 0.76 for the 20°and 40°steering cases, respectively. The corresponding far-field scattering patterns are presented in Fig. 4. It is evident that the proposed method provides substantial sidelobe suppression while maintaining a main beam performance comparable to that of a conventional 2-bit quantized reflectarray. In contrast to [31], where the application of space-time modulation results in a main beam gain reduction of approximately 2 dB or more relative to the 2-

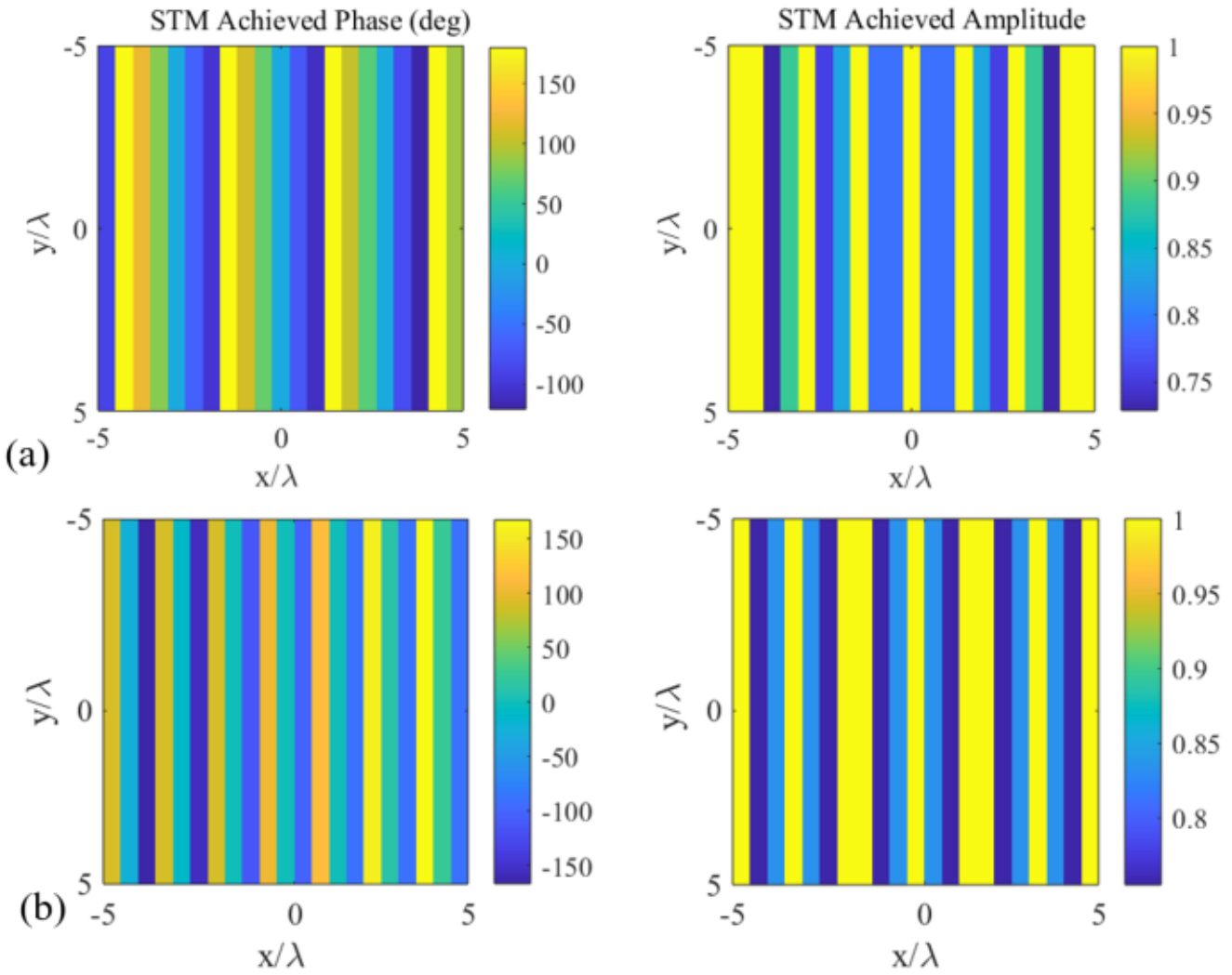

Fig. 3. Aperture phase and amplitude synthesis using space-time modulation (STM) with $L$ = 16, assuming a phase tolerance $\Delta\phi$ of 15º for an aperture size of 10λ × 10λ with inter-element spacing of d = λ/2 and beam steering angles of (a) $\theta$ = 20º, $\phi$ = 0º and (b) $\theta$ = 40º, $\phi$ = 0º.

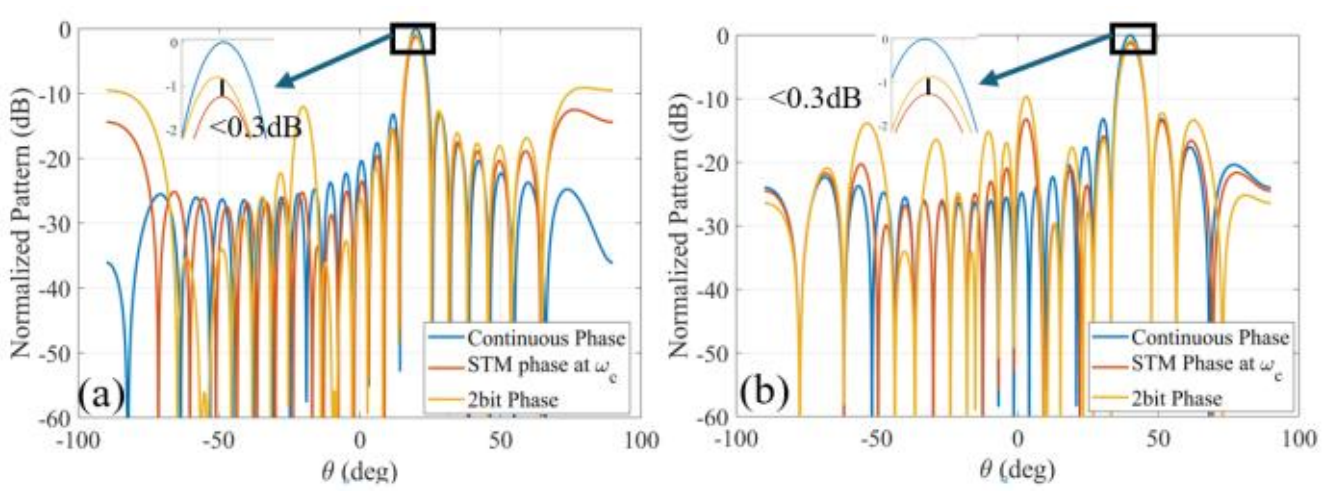

Fig. 4. Normalized radiation pattern from an aperture having size of 10λ × 10λ with inter-element spacing of d = λ/2 and beam steering angles of (a) $\theta$ = 20º, $\phi$ = 0º and (b) $\theta$ = 40º, $\phi$ = 0º.

bit architecture, the proposed approach incurs only a 0.3 dB reduction in main-beam gain. These results demonstrate that the optimized amplitude-phase synthesis effectively mitigates the efficiency degradation typically associated with space-time modulation while simultaneously improving sidelobe performance.

To further quantify the performance degradation associated with space-time modulation, Table I summarizes the main beam gain reduction, pointing error, and the minimum and average synthesized aperture amplitudes as a function of the allowable phase tolerance, ranging from 0°to 40°for a 2-bit control architecture. It can be observed that the gain loss is most pronounced under a strict phase-matching criterion. For a phase tolerance of 0°, the synthesized aperture exhibits amplitudes as low as 0.125, resulting in an average aperture amplitude of only 0.621. This substantial reduction in effective aperture illumination is the primary reason for the degraded beam performance reported in earlier space-time modulated reflectarrays employing equivalent 3-bit and 4-bit phase synthesis [31]. As the phase tolerance is gradually relaxed, higher-amplitude states can be utilized more frequently, leading to improved aperture efficiency and reduced main beam loss. Consequently, for phase tolerances of 5°and above, the gain performance progressively approaches that of the conventional 2-bit quantized reflectarray. At a phase tolerance of 40°, the space-time modulation approach effectively converges to the standard 2-bit quantization case.

Among the investigated cases, a phase tolerance of 15°provides an attractive tradeoff between gain preservation and sidelobe suppression. At this operating point, the proposed method achieves a significant reduction in sidelobe levels while incurring only about 0.3 dB of main-beam gain degradation compared to the conventional 2-bit architecture. Such a compromise is particularly beneficial for high-power directed-energy systems, where 1-2 dB could make a drastic impact, *i.e.* pushing 25-58% more energy into space with controlled sidelobes to avoid damage in the undesired direction. Similarly, in satellite communication systems, sidelobes are tightly constrained by regulatory requirements due to their potential to cause interference with neighboring satellites and services. The proposed approach offers a practical means of achieving lower sidelobe levels without sacrificing aperture efficiency, potentially reducing the required transmit power by up to approximately 25-58% for the same link performance and thereby lowering the size, weight, and power (SWaP) requirements of spaceborne hardware.

From a practical implementation perspective, the proposed 16-segment temporal discretization increases the required control clock rate by a factor of sixteen relative to the modulation frequency, placing greater demands on the biasing network, synchronization circuitry, and digital controller. However, this requirement remains compatible with the nanosecond switching speeds of modern PIN diodes and therefore does not fundamentally limit the achievable beam steering rate. The primary implementation challenges arise from finite diode rise and fall times, timing jitter, switching losses, and parasitic effects in the bias network, which may introduce deviations from the ideal temporal coding sequence and increase undesired harmonic content. Furthermore, higher modulation rates can lead to increased power consumption and electromagnetic interference within the control electronics. Nevertheless, these challenges need to be carefully addressed for the practical implementation of the proposed approach.

TABLE I
STM LOSS FOR STEERING ANGLE OF $\theta = 40^{o}$

| Phase Tolerance ($\Delta\phi$) | Main Beam Loss (dB) | Pointing Error ($\Delta\theta$) deg | Average Amplitude | Minimum Amplitude |
|---|---|---|---|---|
| 0 | -4.136 | 0 | 0.621 | 0.125 |
| 5 | -1.789 | -0.1 | 0.815 | 0.713 |
| 10 | -1.531 | -0.1 | 0.845 | 0.729 |
| 15 | -1.28 | -0.2 | 0.883 | 0.755 |
| 20 | -1.196 | -0.2 | 0.897 | 0.755 |
| 25 | -1.086 | -0.2 | 0.923 | 0.791 |
| 30 | -0.977 | -0.2 | 0.953 | 0.834 |
| 35 | -0.922 | -0.3 | 0.983 | 0.940 |
| 40 | -0.895 | -0.3 | 1.00 | 1.00 |

## V. CONCLUSION

An amplitude-aware synthesis methodology for STM reflect-arrays has been presented to address the main beam gain degradation that can arise when equivalent phase states are selected based solely on phase matching. By leveraging the diverse phase-amplitude states generated through temporal coding, the proposed approach introduces a phase-tolerance-based selection strategy that prioritizes high-amplitude reflection coefficients while preserving the desired aperture phase distribution. The results demonstrate that the additional degrees of freedom offered by STM can be utilized not only to enhance phase resolution but also to improve aperture efficiency through informed state selection. Consequently, the proposed technique enables improved beamforming performance and reduced quantization-induced sidelobes, offering a practical framework for realizing efficient, electronically reconfigurable reflect-arrays based on low-complexity 2-bit unit-cell architectures.